\documentstyle[12pt,epsf,epsfig]{article}
\def\beq{\begin{equation}}  
\def\eeq{\end{equation}}
\def\lsim{\mathrel{\rlap{\lower3pt\hbox{\hskip0pt$\sim$}}
    \raise1pt\hbox{$<$}}}         
\def\gsim{\mathrel{\rlap{\lower4pt\hbox{\hskip1pt$\sim$}}
    \raise1pt\hbox{$>$}}}         
\def\simlt{\mathrel{\raise.3ex\hbox{$<$\kern-.75em\lower1ex\hbox{$\sim$}}}}
\def\simgt{\mathrel{\raise.3ex\hbox{$>$\kern-.75em\lower1ex\hbox{$\sim$}}}}
\def\ga{\mathrel{\raise.3ex\hbox{$>$\kern-.75em\lower1ex\hbox{$\sim$}}}}
\def\la{\mathrel{\raise.3ex\hbox{$<$\kern-.75em\lower1ex\hbox{$\sim$}}}}

\def\PR{{\it Phys.Rev.} }

\def\ohsq{\Omega_{\chi} h^2}
\def\m12{m_{1\!/2}}
\def\mst{m_{\tilde\tau_1}}
\def\stau{\tilde \tau}

\def\st{{\widetilde \tau}_{\scriptscriptstyle\rm 1}}

\def\gev{{\rm \, Ge\kern-0.125em V}}

\def\nnl{\hfill\nonumber\\}
\def\mz{m_{\ss Z}}
\def\ss{\scriptscriptstyle}
\def\nl{\hfill\nonumber\\&&}
\def\ttbt{\tan^2 \beta}

\begin{document}
\begin{titlepage}

\begin{flushright}
hep-ph/0211064 \\
TPI--MINN--02/45\\
UMN--TH--2117/02 \\
November 2002
\end{flushright}
\begin{center}
\baselineskip25pt

\vspace{1cm}

{\Large\bf CDM in Supersymmetric Models\footnote{Summary of an invited
talk at the 10th International Conference on Supersymmetry and
Unification of Fundamental Interactions (SUSY02) held at DESY, Hamburg,
Germany, June 17-22, 2002. }}

\vspace{1cm}

{\sc
 Keith A. Olive$^{a}$}
\vspace{0.3cm}

$^a${\it Theoretical Physics Institute, School of Physics and Astronomy,
\\ University of Minnesota, Minneapolis, MN 55455}

\end{center}
\vspace{1cm}
\begin{abstract}
The supersymmetric extension to the Standard Model offers a promising
cold dark matter candidate, the lightest neutralino. I will review the
phenomenological and cosmological constraints on the
supersymmetric parameter space and discuss the prospects for the detection
of this candidate in both accelerator and direct detection searches.

\end{abstract}

\end{titlepage}

\section{Introduction}
It is well known that supersymmetric models with conserved $R$-parity
contain one new stable particle which is a candidate for cold dark matter
(CDM) \cite{EHNOS}. There are very strong constraints, however, forbidding the existence of stable or
long lived particles which are not color and electrically neutral.
Strong and electromagnetically interacting LSPs would become bound with
normal matter forming anomalously heavy isotopes. Indeed, there are very
strong upper limits on the abundances, relative to hydrogen, of nuclear
isotopes~\cite{isotopes},
$n/n_H \la 10^{-15}~~{\rm to}~~10^{-29}
$
for 1 GeV $\la m \la$ 1 TeV. A strongly interacting stable relic is
expected to have an abundance $n/n_H \la 10^{-10}$
with a higher abundance for charged particles. 
There are relatively few supersymmetric candidates which are not colored
and are electrically neutral.  The sneutrino \cite{snu,fos} is one
possibility, but in the MSSM, it has been excluded as a dark matter
candidate by direct \cite{dir} and indirect \cite{indir} searches.  In
fact, one can set an accelerator based limit on the sneutrino mass from
neutrino counting, 
$m_{\tilde\nu}\ga$ 44.7 GeV \cite{EFOS}. In this case, the direct relic
searches in
underground low-background experiments require  
$m_{\tilde\nu}\ga$ 20 TeV~\cite{dir}. Another possibility is the
gravitino which is probably the most difficult to exclude. 
I will concentrate on the remaining possibility in the MSSM, namely the
neutralinos.

 There are four neutralinos, each of which is a  
linear combination of the $R=-1$, neutral fermions \cite{EHNOS}: the wino
$\tilde W^3$, the partner of the
 3rd component of the $SU(2)_L$ gauge boson;
 the bino, $\tilde B$, the partner of the $U(1)_Y$ gauge boson;
 and the two neutral Higgsinos,  $\tilde H_1$ and $\tilde H_2$.
 In general,
neutralinos can  be expressed as a linear combination
\begin{equation}
	\chi = \alpha \tilde B + \beta \tilde W^3 + \gamma \tilde H_1 +
\delta
\tilde H_2
\end{equation}
The solution for the coefficients $\alpha, \beta, \gamma$ and $\delta$
for neutralinos that make up the LSP 
can be found by diagonalizing the mass matrix
\beq
      ({\tilde W}^3, {\tilde B}, {{\tilde H}^0}_1,{{\tilde H}^0}_2 )
  \left( \begin{array}{cccc}
M_2 & 0 & {-g_2 v_1 \over \sqrt{2}} &  {g_2 v_2 \over \sqrt{2}} \\
0 & M_1 & {g_1 v_1 \over \sqrt{2}} & {-g_1 v_2 \over \sqrt{2}} \\
{-g_2 v_1 \over \sqrt{2}} & {g_1 v_1 \over \sqrt{2}} & 0 & -\mu \\
{g_2 v_2 \over \sqrt{2}} & {-g_1 v_2 \over \sqrt{2}} & -\mu & 0 
\end{array} \right) \left( \begin{array}{c} {\tilde W}^3 \\
{\tilde B} \\ {{\tilde H}^0}_1 \\ {{\tilde H}^0}_2 \end{array} \right)
\eeq
where $M_1 (M_2)$ are the soft supersymmetry breaking
 U(1) (SU(2))  gaugino mass terms. $\mu$ is the supersymmetric Higgs
mixing mass parameter and since there are two Higgs doublets in
the MSSM, there are  two vacuum expectation values, $v_1$ and $v_2$. One
combination of these is related to the $Z$ mass, and therefore is not a
free parameter, while the other combination, the ratio of the two vevs,
$\tan \beta$, is free.

The most general version of the MSSM, despite its minimality in particles and
interactions contains well over a hundred new parameters. The study of such a model
would be untenable were it not for some (well motivated) assumptions.
These have to do with the parameters associated with supersymmetry breaking.
It is often assumed that, at some unification scale, all of the gaugino masses
receive a common mass, $m_{1/2}$. The gaugino masses at the weak scale are
determined by running a set of renormalization group equations.
Similarly, one often assumes that all scalars receive a common mass, $m_0$,
at the GUT scale (though one may wish to make an exception for the Higgs
soft masses). These too are run down to the weak scale. The remaining
supersymmetry breaking parameters are the trilinear mass terms, $A_0$,
which I will also assume are unified at the GUT scale,  and the bilinear
mass term
$B$. There are, in addition, two physical CP violating phases which will
not be considered here.

The natural boundary conditions at the GUT scale for the MSSM would
include
$\mu$, the two soft Higgs masses ($m_1$ and $m_2$) and $B$ in addition to
$m_{1/2}$,
$m_0$, and $A_0$. In this case, by running the RGEs down to a low energy
scale, one would predict the values of $M_Z$, $\tan \beta$, and the Higgs
pseudoscalar mass, $m_A$ (in addition to all of the sparticle masses).
Since $M_Z$ is known, it is more useful to analyze supersymmetric models
where $M_Z$ is input rather than output.  It is also common to treat
$\tan \beta$ as an input parameter. This can be done at the expense of 
shifting $\mu$ (up to a sign) and $B$ from inputs to outputs. 
When the supersymmetry breaking Higgs soft masses are also unified at the
GUT scale (and take the common value $m_0$), the model is often referred
to as the constrained MSSM or CMSSM. Once these parameters are set, the
entire spectrum of sparticle masses at the weak scale can be calculated.

\section{The Relic Density}

The relic abundance of LSP's is 
determined by solving
the Boltzmann
 equation for the LSP number density in an expanding Universe.
 The technique \cite{wso} used is similar to that for computing
 the relic abundance of massive neutrinos \cite{lw}.
The relic density depends on additional parameters in the MSSM beyond $m_{1/2},
\mu$, and $\tan \beta$. These include the sfermion masses, $m_{\tilde
f}$, as well as  $m_A$, all derived from $m_0$, $A_0$, and
$m_{1/2}$. To determine the relic density it is necessary to obtain the
general annihilation cross-section for neutralinos.  In much of the
parameter space of interest, the LSP is a bino and the annihilation
proceeds mainly through sfermion exchange. Because of the p-wave
suppression associated with Majorana fermions, the s-wave part of the
annihilation cross-section is suppressed by the outgoing fermion masses. 
This means that it is necessary to expand the cross-section to include
p-wave corrections which can be expressed as a term proportional to the
temperature if neutralinos are in equilibrium. Unless the neutralino mass
happens to lie near near a pole, such as $m_\chi \simeq$
$m_Z/2$ or $m_h/2$, in which case there are large contributions to the
annihilation through direct $s$-channel resonance exchange, the dominant
contribution to
the $\tilde{B} \tilde{B}$ annihilation cross section comes from crossed
$t$-channel sfermion exchange.

Annihilations in the early
Universe continue until the annihilation rate
$\Gamma
\simeq \sigma v n_\chi$ drops below the expansion rate given by the Hubble parameter,
$H$.  For particles which annihilate through approximate weak scale interactions, this
occurs when $T \sim m_\chi /20$. Subsequently, the relic density of neutralinos is
fixed relative to the number of relativistic particles. 
As noted above, the number density of neutralinos is tracked by a
Boltzmann-like equation,
\beq
{dn \over dt} = -3{{\dot R} \over R} n - \langle \sigma v \rangle (n^2 -
n_0^2)
\label{rate}
\eeq
where $n_0$ is the equilibrium number density of neutralinos.
By defining the quantity $f = n/T^3$, we can rewrite this equation in
terms of $x = T/m_\chi$, as
\beq
{df \over dx} = m_\chi \left( {4 \over 45} \pi^3 G_N N \right)^{-1/2}
\langle \sigma v \rangle (f^2 - f_0^2)
\label{rate2}
\eeq
where $N$ is the number of relativistic degrees of freedom.
The solution to this equation at late times (small $x$) yields a constant value of
$f$, so that $n \propto T^3$. 
The final relic density expressed as a fraction of the critical energy density 
can be written as \cite{EHNOS}
\beq
\Omega_\chi h^2 \simeq 1.9 \times 10^{-11} \left({T_\chi \over
T_\gamma}\right)^3 N_f^{1/2} \left({{\rm GeV} \over ax_f + {1\over 2} b
x_f^2}\right)
\label{relic}
\eeq 
where $(T_\chi/T_\gamma)^3$ accounts for the subsequent reheating of the
photon temperature with respect to $\chi$, due to the annihilations of
particles with mass $m < x_f m_\chi$ \cite{oss}. The subscript $f$ refers to
values at freeze-out, i.e., when annihilations cease. The coefficients $a$ and $b$
are related to the partial wave expansion of the cross-section, $\sigma v = a + b x +
\dots $. Eq. (\ref{relic} ) results in a very good approximation to the relic density
expect near s-channel annihilation poles,  thresholds and in regions where the LSP is
nearly degenerate with the next lightest supersymmetric particle
\cite{gs}.

The preferred range of the relic LSP density is provided by data on
the cosmic microwave background (CMB), which have recently been used to
obtain the following 95\% confidence range: $\Omega_{\rm 
CDM} h^2 = 0.12 \pm 0.04$~\cite{MS}. Values much smaller than
$\Omega_{\rm CDM} h^2 = 0.10$ seem to be disfavoured by earlier analyses
of structure formation in the CDM framework, so we restrict our attention
to $\Omega_{\rm CDM} h^2 > 0.1$. However, one should note that the LSP
may not constitute all the CDM, in which case
$\Omega_{\rm LSP}$ could be reduced below this value. On the upper side,
it is preferable to remain very conservative, in particular because the
upper limit on
$\Omega_{\rm LSP}$ sets the upper limit for the sparticle mass scale.
Here, 
$\Omega_{\rm CDM} h^2 < 0.3$ is used, while being aware that the lower
part of this range currently appears the most plausible.

\begin{figure}
\hspace*{1.3in}
\epsfig{file=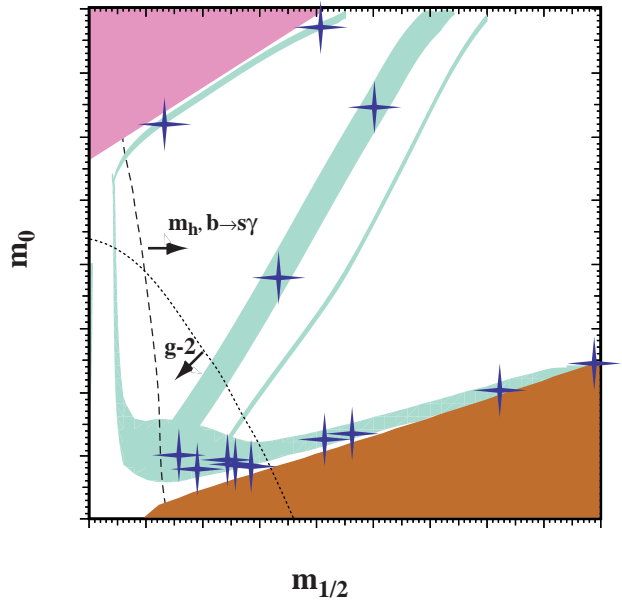,height=3.2in}
\caption{\label{fig:rough}
{\it 
Qualitative overview of 
a generic $(m_{1/2}, m_0)$ plane.  The light (turquoise) shaded area is
the cosmologically preferred region with \protect\mbox{$0.1\leq\ohsq\leq
0.3$}.  In the
dark (brick red) shaded region at bottom right, the LSP is the charged
${\tilde \tau}_1$, so this region is excluded. Electroweak symmetry
breaking is not possible in the dark (pink) shaded region at top left. The
LEP experimental constraints, in particular that on $m_h$, and
measurements of $b \rightarrow s \gamma$ exert pressure from the left
side. The BNL E821 measurement of $g_\mu - 2$ favours relatively low
values of $m_0$ and $m_{1/2}$ for $\mu > 0$. 
}}
\end{figure}

Fig.~\ref{fig:rough} illustrates qualitatively the regions of the
$(m_{1/2}, m_0)$ plane favoured by LEP limits, particularly on $m_h$, $b
\rightarrow s \gamma$ and cosmology. Electroweak symmetry breaking is not
possible in the dark-shaded triangular region in the top left corner, and
the lightest supersymmetric particle would be charged in the bottom right
dark-shaded triangular region. The experimental constraints on $m_h$ and
$b \rightarrow s \gamma$ exert pressures from the left, as indicated,
which depend on the value of $\tan \beta$ and the sign of $\mu$. The
indication of a deviation from the Standard Model in $g_\mu - 2$
disfavours $\mu < 0$ and large values of $m_0$ and $m_{1/2}$ for $\mu >
0$. The region where
$\ohsq$ falls within the preferred range is indicated in light shading,
its exact shape being dependent on the value of $\tan\beta$.  As
discussed later in more detail, in addition to the `bulk' region at low
$m_0$ and $m_{1/2}$, there are a number of regions at large values
of $m_{1/2}$ and/or $m_0$ where the relic density is still compatible
with the cosmological constraints. At large values of $m_{1/2}$, the
lighter stau, becomes nearly degenerate with the neutralino and
coannihilations between these particles must be taken into
account \cite{EFOSi,glp}.  For non-zero values of $A_0$, there are new
regions  for which $\chi-{\tilde t}$  coannihilations are
important \cite{stopco}. At large $\tan \beta$, as one increases
$m_{1/2}$, the pseudo-scalar mass, $m_A$ begins to drop so that there is 
a wide funnel-like region (at all values of $m_0$) such that $2m_\chi
\approx m_A$ and s-channel annihilations become
important \cite{funnel,EFGOSi}. Finally, there is a region at very high
$m_0$ where the value of $\mu$ begins to fall and the LSP becomes more
Higgsino-like.  This is known as the `focus point'
region \cite{Feng:2000gh}. Also shown are the position of several
benchmark points chosen for detailed phenomenological study
\cite{benchmark}.

\begin{figure}
\hspace*{1.3in}
\begin{minipage}{8in}
\epsfig{file=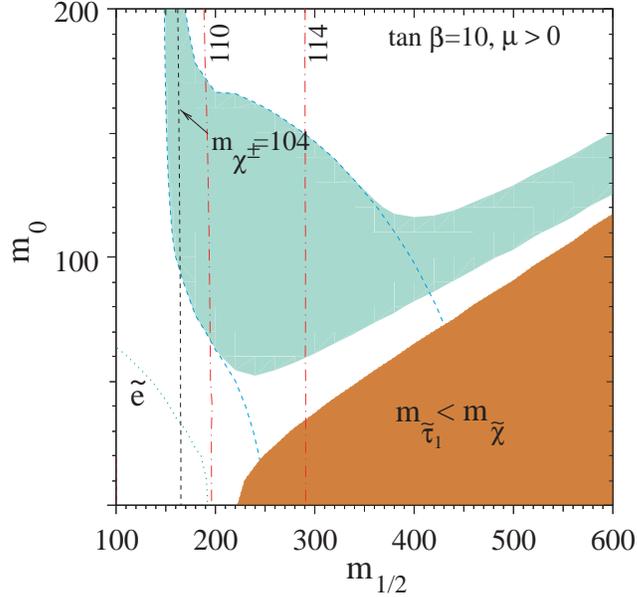,height=3.2in} \hfill
\end{minipage}
\caption{\label{fig:sm}
{\it The light-shaded `bulk' area is the cosmologically preferred 
region with \protect\mbox{$0.1\leq\ohsq\leq 0.3$}.   The light dashed lines
show the location  of the cosmologically preferred region  if one
ignores  coannihilations with the light sleptons.  
In the dark shaded region in the bottom right, the LSP is
the ${\tilde
\tau}_1$, leading to an unacceptable abundance
of charged dark matter.  Also shown is the isomass
contour $m_{\chi^\pm} = 104$~GeV and $m_h = 110,114$~GeV,
as well as an indication of the slepton bound from
LEP.  }}
\end{figure}

Let us first focus on the `bulk' region in the CMSSM for $\tan \beta =
10$ and $\mu > 0$ shown in Fig. \ref{fig:sm} \cite{EFOSi}.  The light
shaded region correspond to
\mbox{$0.1<\ohsq<0.3$}.  The dark shaded region has $m_{\st}< m_\chi$
and is excluded. The light dashed contours
indicate the corresponding region in $\ohsq$ if one ignores the effect of
coannihilations.  Neglecting coannihilations, one would find an upper
bound of $\sim450\gev$ on $\m12$, corresponding to an upper bound of
roughly $200\gev$ on $m_{\tilde B}$.

Coannihilations are important when there are several
particle species $i$, with different masses, and
each with its own number density $n_i$ and 
equilibrium number density $n_{0,i}$.
In this case \cite{gs}, the rate equation (\ref{rate}) still applies,
provided $n$ is
interpreted as the total number density, 
\beq
n \equiv \sum_i n_i \;,
\label{n}
\eeq
$n_0$ as the total equilibrium number density, 
\beq
n_0 \equiv  \sum_i n_{0,i} \;,
\label{neq}
\eeq
and the effective annihilation cross section as
\beq
\langle\sigma_{\rm eff} v_{\rm rel}\rangle \equiv
\sum_{ij}{ n_{0,i} n_{0,j} \over n_0^2}
\langle\sigma_{ij} v_{\rm rel}\rangle \;.
\label{sv2}
\eeq
In eq.~(\ref{rate2}),  $m_\chi$ is now understood as the mass of the
lightest sparticle under consideration.

 Note that this implies that the
ratio of relic densities computed with and without coannihilations is, 
roughly,
\begin{equation}
  \label{eq:R}
 R\equiv{\Omega^0\over\Omega} 
 \approx \left({\hat\sigma_{\rm eff}\over\hat\sigma_0}\right)
\left({x_{\!\ss f}\over x_{\!\ss f}^{0}}\right),
\end{equation}
where $\hat\sigma\equiv a + b x/2$ and sub- and superscripts 0 denote
quantities computed ignoring coannihilations.  The ratio ${x_{\!\ss
    f}^0 / x_{\!\ss f}}\approx 1+x_{\!\ss f}^0 \ln (g_{\rm
  eff}\sigma_{\rm eff}/g_1\sigma_0)$, where 
$g_{\rm eff}\equiv\sum_i g_i (m_i/m_1)^{3/2}e^{-(m_i-m_1)/T}$.
For the case of three degenerate slepton NLSPs, $g_{\rm eff}=\sum_i g_i =8$ and 
${x_{\!\ss f}^0 / x_{\!\ss f}}\approx 1.2$.

The effect of coannihilations is
to create an allowed band about 25-50 $\gev$ wide in $m_0$ for $\m12 \la
1400\gev$, which tracks above the $\mst=m_\chi$ contour.  Along the
line $\mst= m_\chi$, $R\approx10$, from (\ref{eq:R}) \cite{EFOSi}.  As
$m_0$ increases,
the mass difference increases and the slepton contribution to
$\hat\sigma_{\rm  eff}$ falls,  and the relic density rises
abruptly.

\section{The CMSSM with Universal Higgs masses}

\begin{figure}
\begin{minipage}{8in}
\epsfig{file=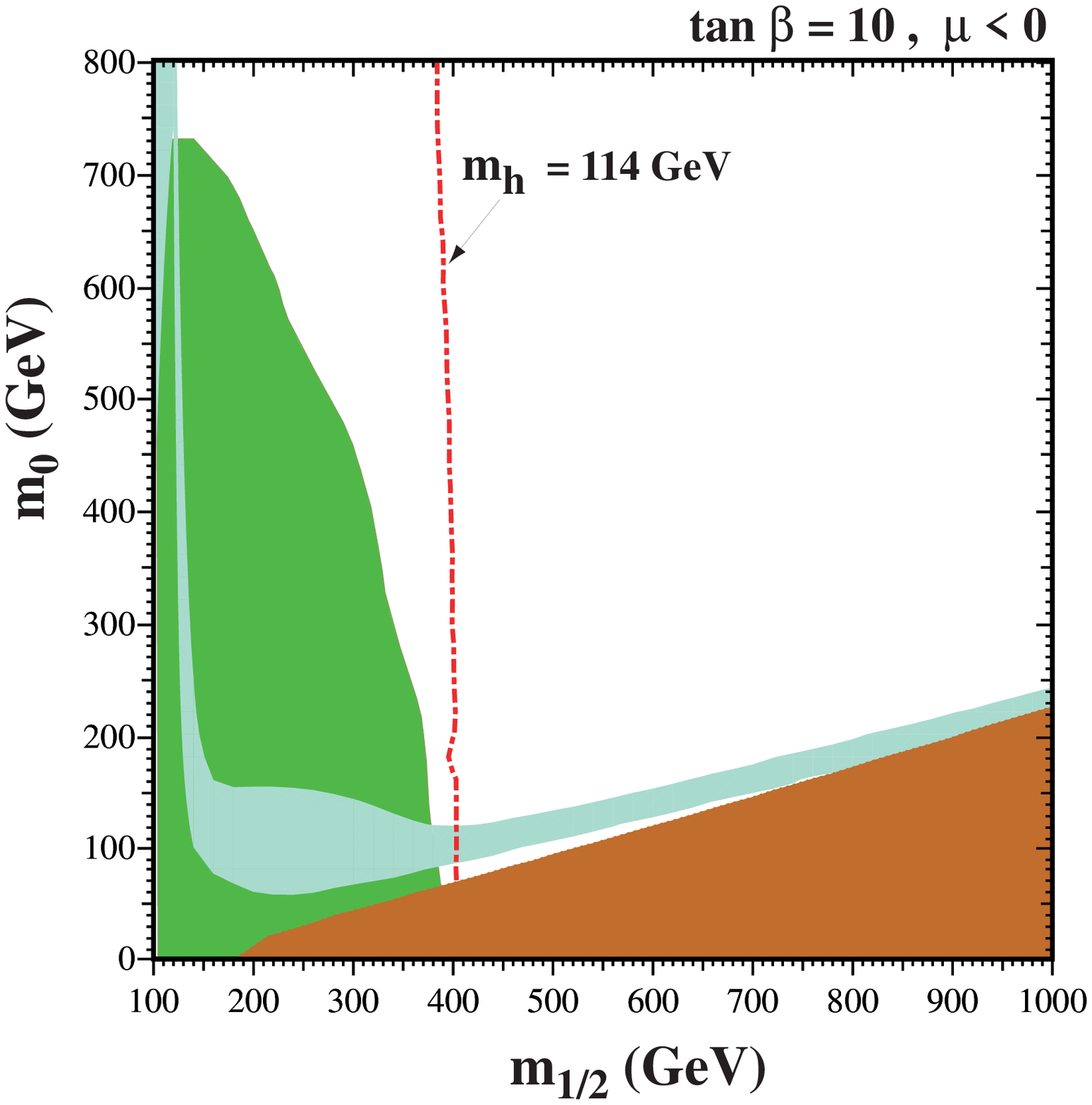,height=3.2in}
\epsfig{file=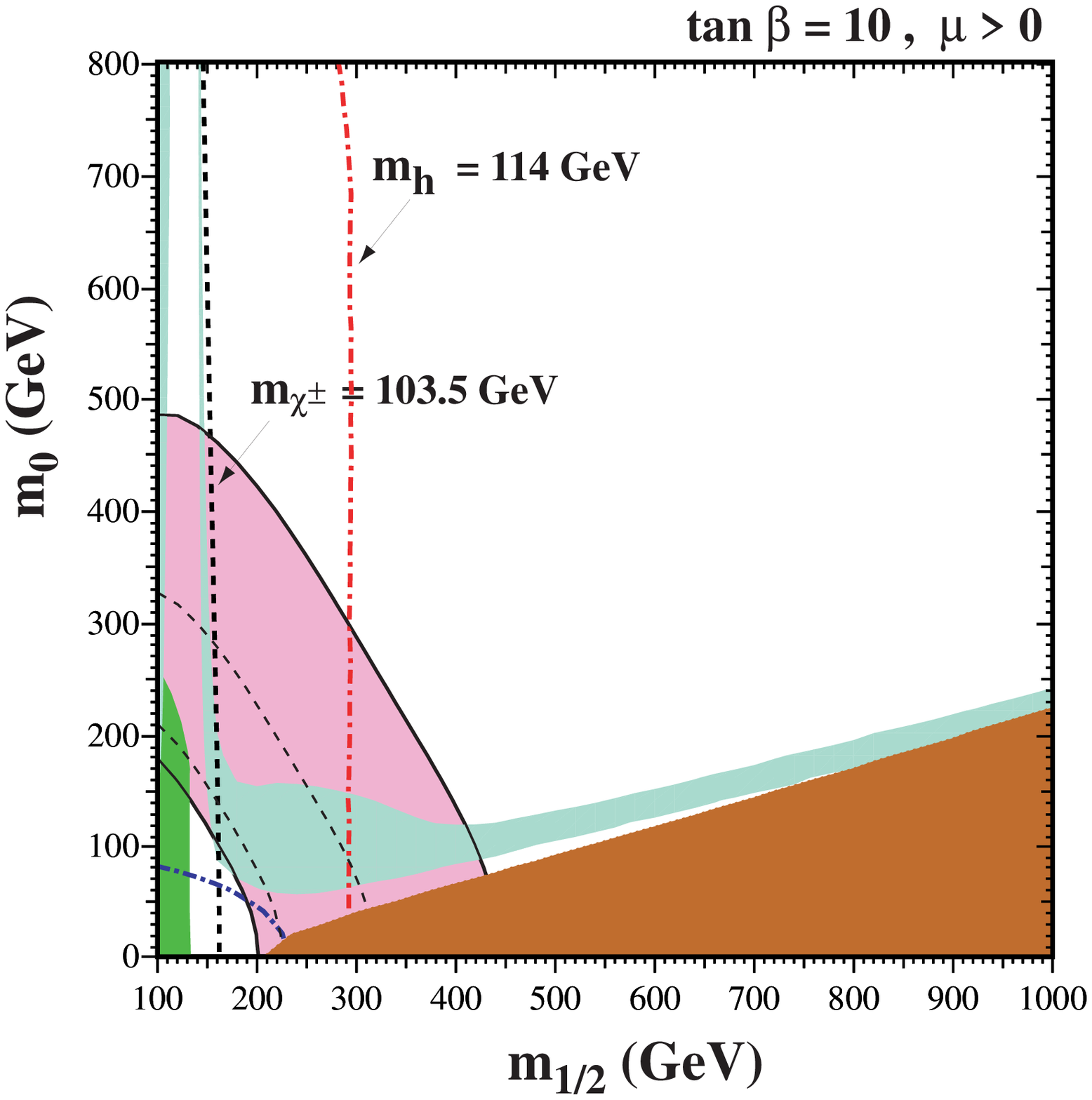,height=3.2in}
\end{minipage}
\caption{\label{fig:UHM}
{\it 
The $(m_{1/2}, m_0)$ planes for  (a) $\tan \beta = 10$ and $\mu < 0$, (b)
$\tan \beta = 10$ and $\mu > 0$, 
assuming $A_0 = 0, m_t = 175$~GeV and
$m_b(m_b)^{\overline {MS}}_{SM} = 4.25$~GeV. The near-vertical (red)
dot-dashed lines are the contours $m_h = 114$~GeV as calculated using
{\tt FeynHiggs}~\cite{FeynHiggs}, and the near-vertical (black) dashed
line is the contour $m_{\chi^\pm} = 103.5$~GeV, shown only in (b). Also
shown in (b) by the dot-dashed curve in the lower left is the corner
excluded by the LEP bound of $m_{\tilde e} > 99$ GeV. The medium (dark
green) shaded region is excluded by $b \to s
\gamma$, and the light (turquoise) shaded area is the cosmologically
preferred regions with \protect\mbox{$0.1\leq\ohsq\leq 0.3$}. In the dark
(brick red) shaded region, the LSP is the charged ${\tilde \tau}_1$. The
region allowed by the E821 measurement of $a_\mu$ at the 2-$\sigma$
level, is shaded (pink) and bounded by solid black lines, with dashed
lines indicating the 1-$\sigma$ ranges.}}
\end{figure}

A larger view of the $\tan \beta = 10$ parameter plane is shown in Fig.
\ref{fig:UHM} \cite{EFGO,EFGOSi,benchmark,EOS2,eos3}.  Included here are
the most important phenomenological constraints  (shown schematically in
Figure~\ref{fig:rough}).  These include the constraints on the MSSM
parameter space that are provided by direct sparticle searches at LEP,
including that on the lightest chargino $\chi^\pm$: $m_{\chi^\pm} \ga$
103.5 GeV~\cite{LEPsusy}, and that on the selectron $\tilde e$: $m_{\tilde
e}\ga$ 99 GeV \cite{LEPSUSYWG_0101}. Another important constraint is
provided by the LEP lower limit on the Higgs mass: $m_H > 114.4$~GeV
\cite{LEPHiggs} in the Standard Model. Since $m_h$ is sensitive to
sparticle masses, particularly
$m_{\tilde t}$, via loop corrections,
the Higgs limit also imposes important constraints on the
CMSSM parameters, principally $m_{1/2}$.

 The constraint imposed by
measurements of $b\rightarrow s\gamma$~\cite{bsg} also excludes small
values of
$m_{1/2}$. These measurements agree with the Standard Model, and
therefore provide bounds on MSSM particles,  such as the chargino and
charged Higgs masses, in particular. Typically, the $b\rightarrow s\gamma$
constraint is more important for $\mu < 0$, but it is also relevant for
$\mu > 0$,  particularly when $\tan\beta$ is large. 

The latest value of the anomalous magnetic
moment of the muon reported~\cite{newBNL} by the BNL E821 experiment is
also taken into account. The world average of $a_\mu\equiv {1\over 2}
(g_\mu -2)$ now deviates by
$(33.9 \pm 11.2) \times 10^{-10}$ from the Standard Model calculation
of~\cite{Davier} using $e^+ e^-$ data, and by $(17 \pm 11) \times
10^{-10}$ from the Standard Model calculation of~\cite{Davier} based on
$\tau$ decay data. Other recent analyses of the $e^+ e^-$ data yield
similar results. On some of the subsequent plots,  the formal
2-$\sigma$ range $11.5 \times 10^{-10} < \delta a_\mu < 56.3 \times
10^{-10}$ is displayed. As one can see, the region preferred by $g-2$
overlaps very nicely with the `bulk' region for $\tan \beta = 10$ and
$\mu > 0$. 

\begin{figure}[hbtp]
	\centering 
\begin{minipage}{8in}
\epsfig{file=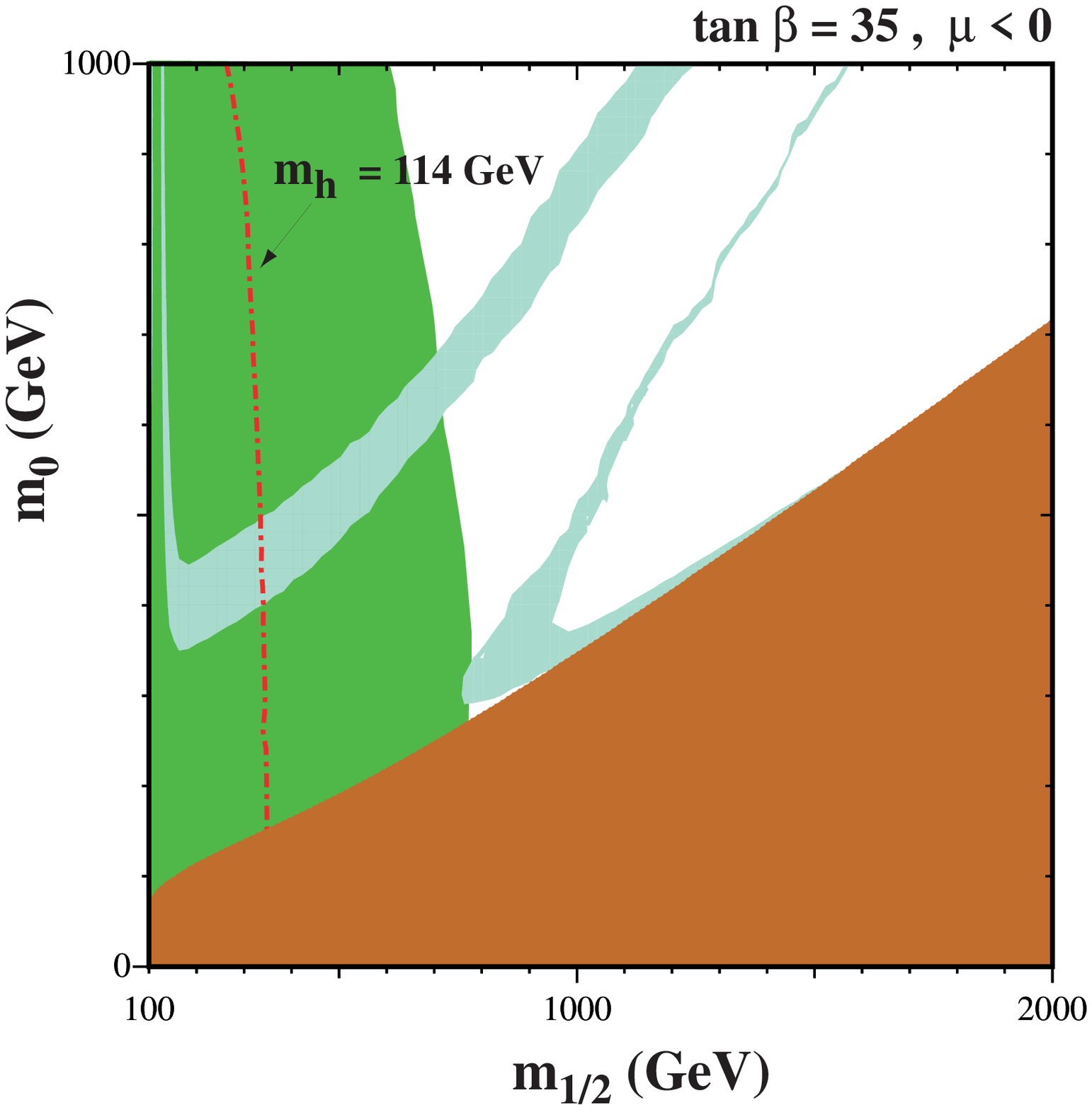,height=3.2in}
\epsfig{file=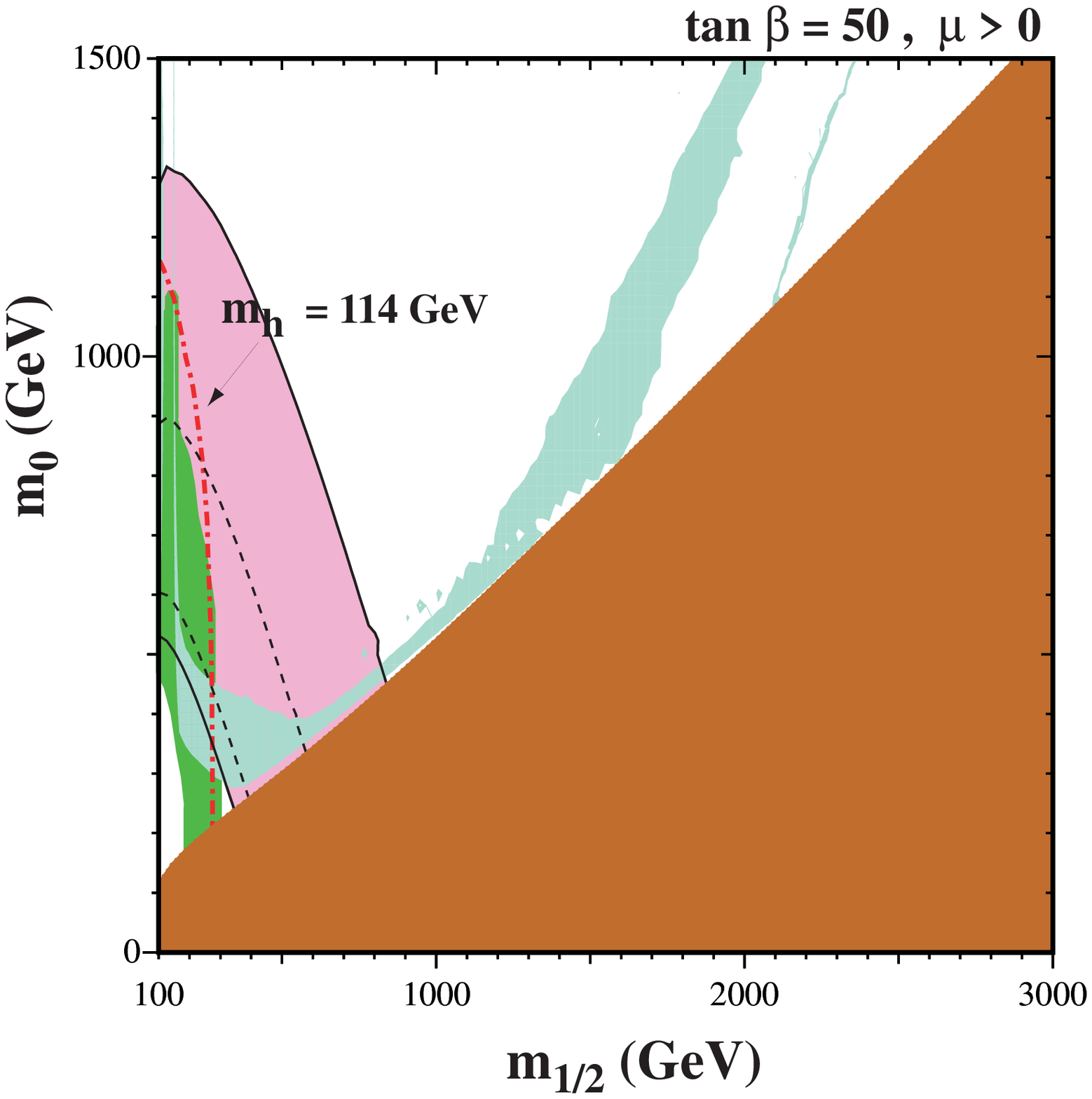,height=3.2in}
\end{minipage}
	\caption{\it As in Fig. \protect\ref{fig:UHM} for (a) $\tan \beta = 35$
and
$\mu < 0$ and (b) $\tan \beta = 50$ and $\mu > 0$ }
	\label{rd2c50}
\end{figure}

As noted above, another
mechanism for extending the allowed CMSSM region to large
$m_\chi$ is rapid annihilation via a direct-channel pole when $m_\chi
\sim {1\over 2} m_{A}$~\cite{funnel,EFGOSi}. Since the heavy scalar and
pseudoscalar Higgs masses decrease as  
$\tan \beta$ increases, eventually  $ 2 m_\chi \simeq  m_A$ yielding a
`funnel' extending to large
$m_{1/2}$ and
$m_0$ at large
$\tan\beta$, as seen in Fig.~\ref{rd2c50}.
The difficulty and necessary care involved in calculations at large 
$\tan \beta$ were discussed in \cite{EFGOSi}. For related CMSSM
calculations see \cite{otherOmega}.

\begin{figure}[hbtp]
	\centering 
\begin{minipage}{8in}
\epsfig{file=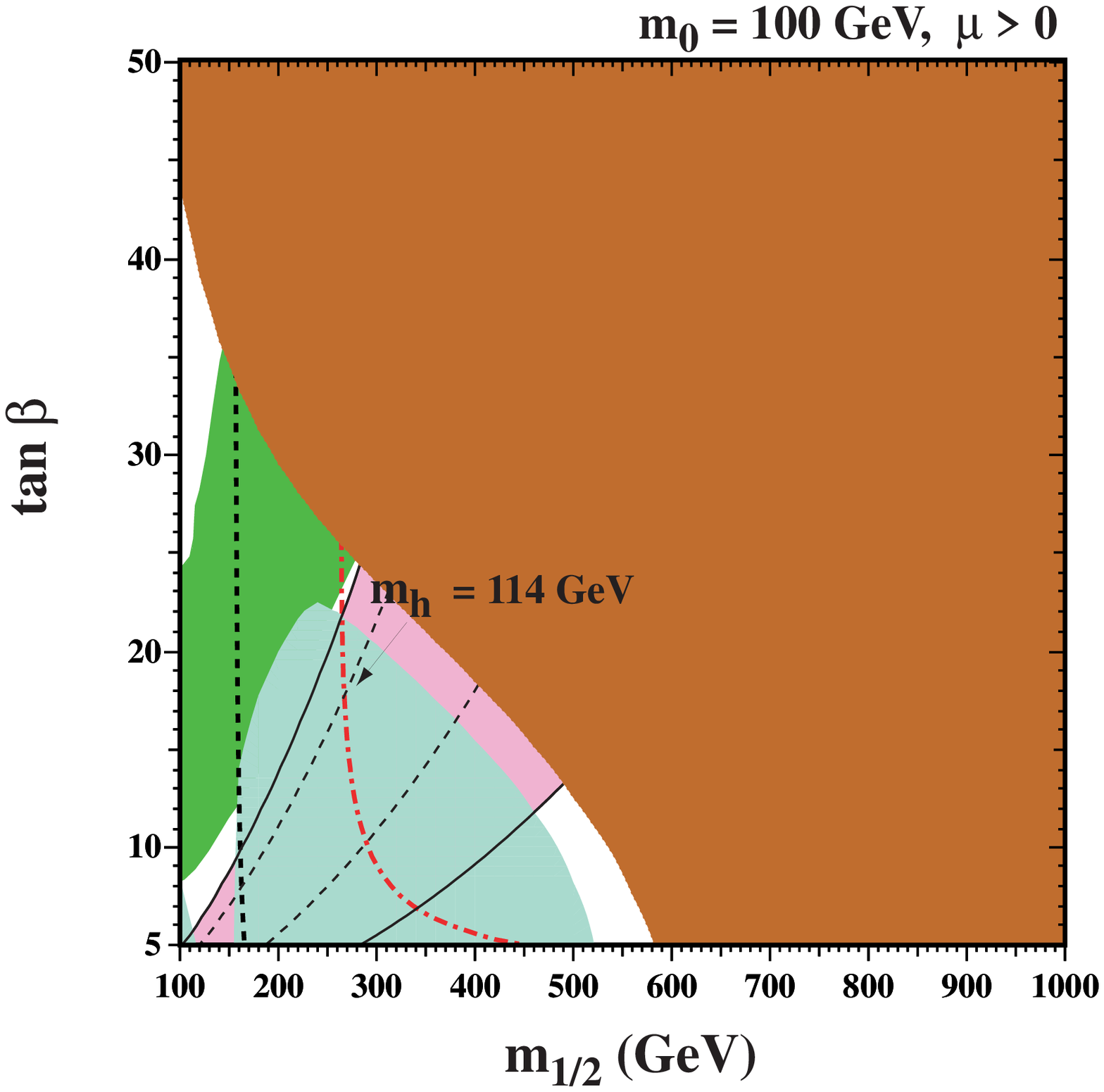,height=3.2in}
\epsfig{file=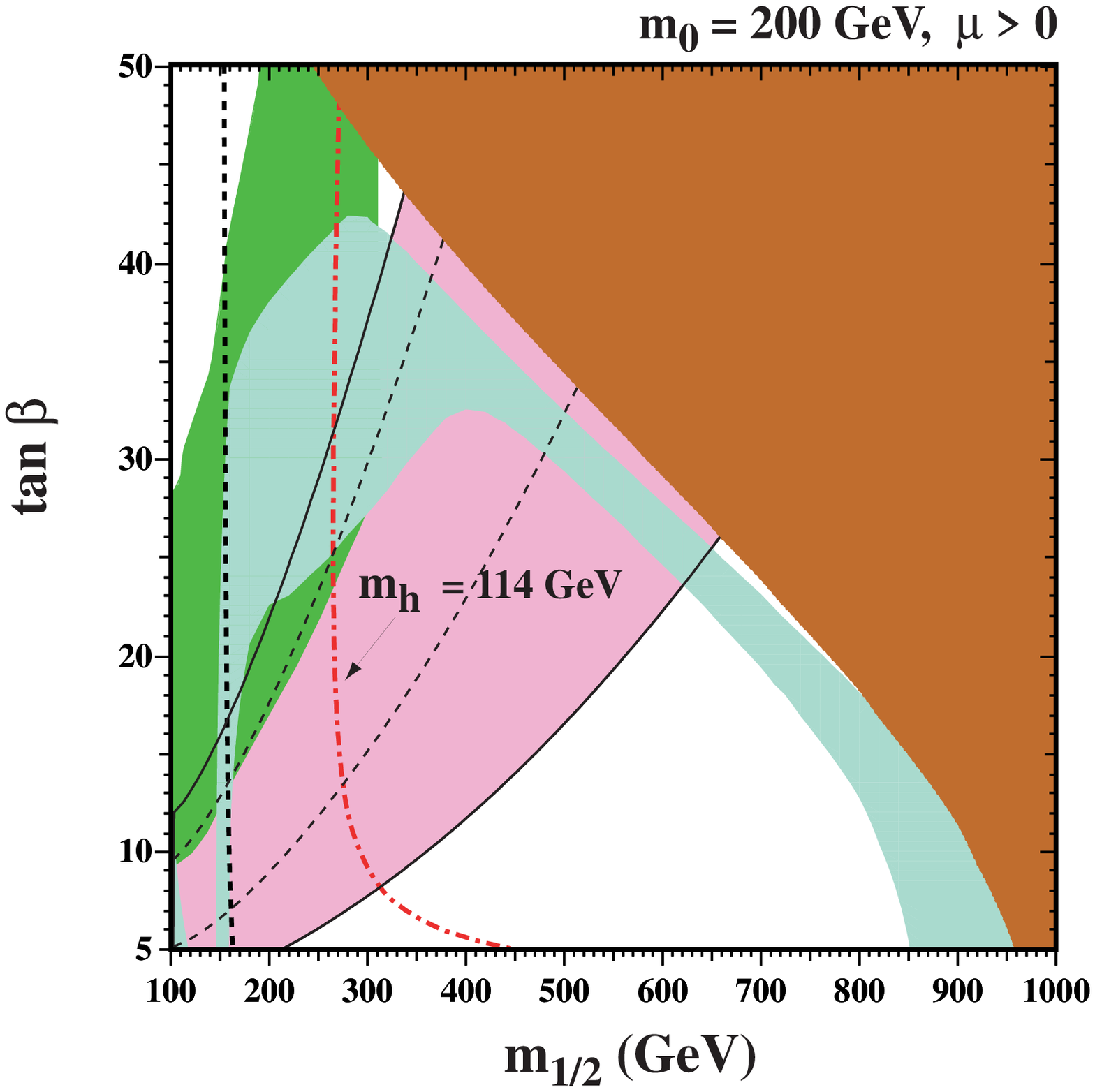,height=3.2in}
\end{minipage}
	\caption{\it The $\tan \beta$ - $m_{1/2}$ plane for (a) $m_0 = 100$ GeV
and (b) $m_0 = 200$ GeV both for $\mu > 0 $. All shading and line types
are as in the previous figures.}
	\label{newones}
\end{figure}

Another way to see the effects of increasing $\tan \beta$ is plot
look at the $\tan \beta - m_{1/2}$ plane for fixed $m_0$.
This is shown in Fig. \ref{newones}. As one can see, there is a charged
$\st$ LSP in the upper right corner for both values of $m_0$, though the 
area is reduced when $m_0$ is increased as is expected. One also sees the
cosmological region moving up in $\tan \beta$ as $m_0$ is increased.  In
b), one also sees the well defined coannihilation tail. Here, the
importance of $b \to s \gamma$ at large $\tan \beta$ is clearly displayed.

\section{The CMSSM with Non-Universal Higgs masses}

If one relaxes the unification condition for the two soft Higgs mass,
$m_1$ and $m_2$, one could decide to make $\mu$ and $m_A$ inputs
and instead making the Higgs soft masses outputs. This is referred to as
the CMSSM with non-universal Higgs masses or NUHM \cite{nonu,EFGOS,eos3}.
The attractive feature of radiative symmetry breaking in the CMSSM is
maintained by using the electroweak minimization conditions to solve for 
$m_1$ and $m_2$, given values of $\mu$ and $m_A$ at the weak scale.

The electroweak symmetry breaking conditions may be written in the form:
\begin{equation}
m_A^2 (Q) = m_1^2(Q) + m_2^2(Q) + 2 \mu^2(Q) + \Delta_A(Q)
\end{equation}
and
\begin{equation}
\mu^2 = \frac{m_1^2 - m_2^2 \tan^2 \beta + \frac{1}{2} \mz^2 (1 - \tan^2 \beta)
+ \Delta_\mu^{(1)}}{\tan^2 \beta - 1 + \Delta_\mu^{(2)}},
\end{equation}
where $\Delta_A$ and $\Delta_\mu^{(1,2)}$ are loop
 corrections~\cite{Barger:1993gh,deBoer:1994he,Carena:2001fw} and
$m_{1,2} \equiv m_{1,2}(\mz)$.  
The known radiative
corrections~\cite{Barger:1993gh,IL,Martin:1993zk} 
$c_1, c_2$ and $c_\mu$ relating the values of the
NUHM parameters at $Q$ to their values at $\mz$ are incorporated: 
\begin{eqnarray}
m_1^2(Q) &=& m_1^2 + c_1 \nnl
m_2^2(Q) &=& m_2^2 + c_2 \nnl
\mu^2(Q) &=& \mu^2 + c_\mu \, .
\end{eqnarray}
Solving for $m^2_1$ and $m^2_2$, one has
\begin{eqnarray}
m_1^2(1+ \tan^2 \beta) &=& m_A^2(Q) \tan^2 \beta - \mu^2 (\tan^2 \beta + 1 -
\Delta_\mu^{(2)} ) 
- (c_1 + c_2 + 2 c_\mu) \ttbt \nl - \Delta_A(Q) \ttbt 
- \frac{1}{2} \mz^2 (1 - \ttbt) - \Delta_\mu^{(1)} 
\label{m1}
\end{eqnarray}
and 
\begin{eqnarray}
m_2^2(1+ \tan^2 \beta) &=& m_A^2(Q) - \mu^2 (\tan^2 \beta + 1 +
\Delta_\mu^{(2)} )
- (c_1 + c_2 + 2 c_\mu) \nl
- \Delta_A(Q) + \frac{1}{2} \mz^2 (1 - \ttbt) + \Delta_\mu^{(1)},
\label{m2}
\end{eqnarray}
which are used to perform the numerical calculations \cite{eos3}. 

It can be seen from (\ref{m1}) and (\ref{m2}) that, if $m_A$ is too small
or $\mu$ is too large, then $m_1^2$ and/or $m_2^2$ can become negative and
large. This could lead to $m_1^2(M_X) + \mu^2(M_X) < 0$ and/or $m_2^2(M_X)
+ \mu^2(M_X) < 0$, thus triggering
electroweak symmetry breaking at the GUT scale. The requirement that
electroweak symmetry breaking occurs far below the GUT scale forces one to
impose the conditions $m_1^2(M_X)+ \mu(M_X), m_2^2(M_X)+ \mu(M_X) > 0$ as
extra constraints, which we call the GUT
stability constraint~\footnote{For a different point of view, however,
see~\cite{fors}.}.

\begin{figure}[hbtp]
	\centering 
\begin{minipage}{8in}
\epsfig{file=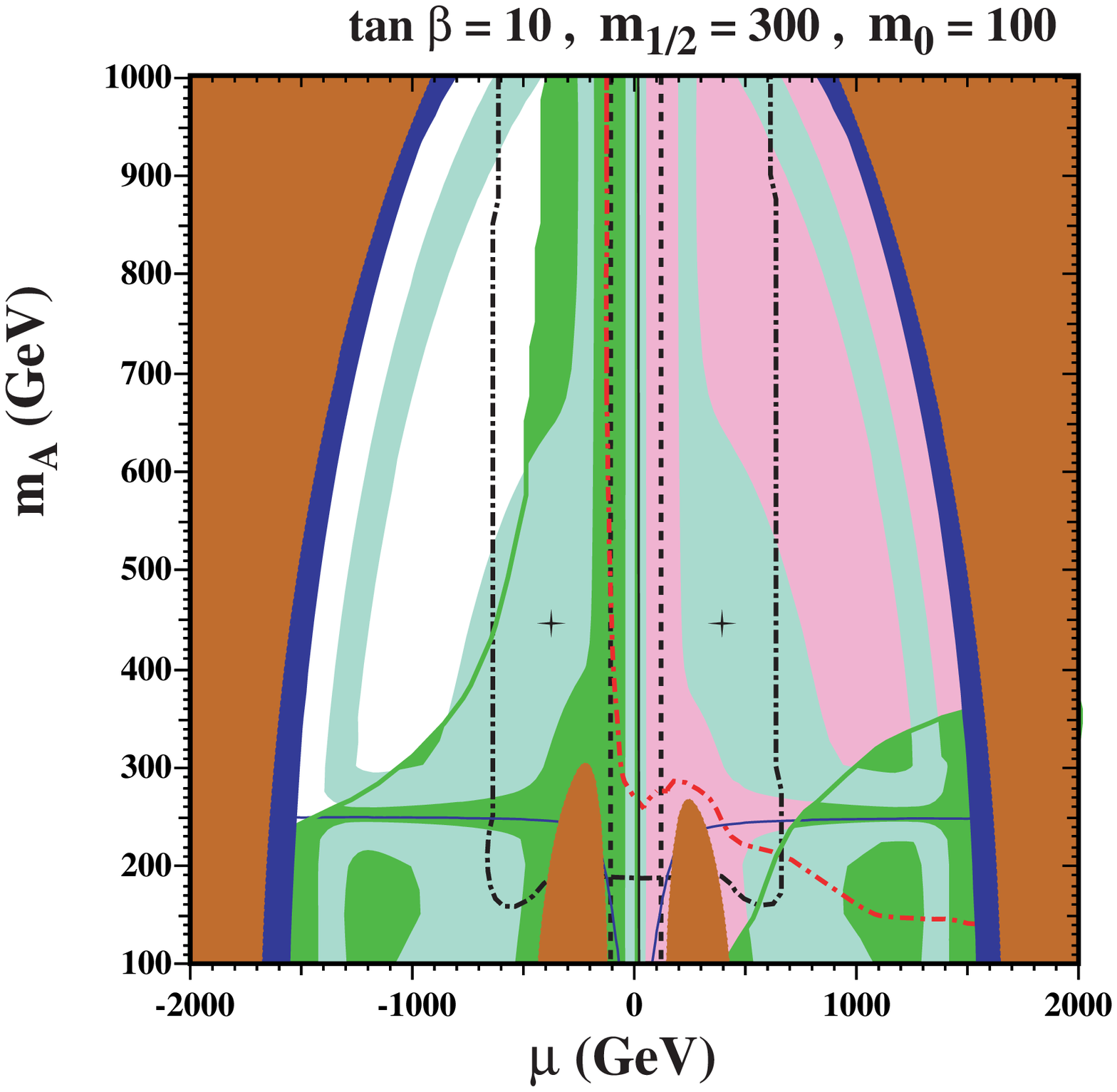,height=3.2in}
\epsfig{file=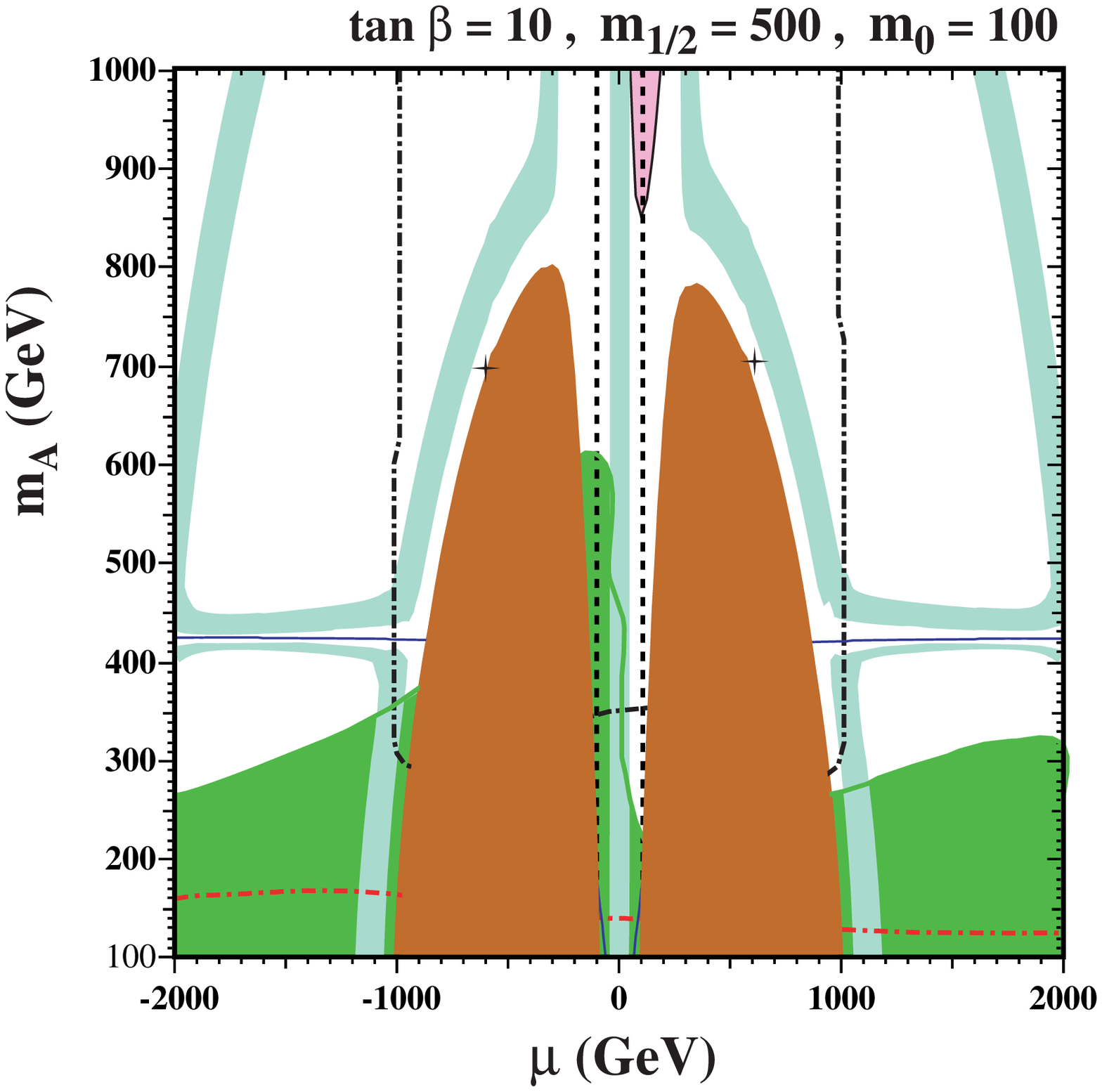,height=3.2in}
\end{minipage}
\caption{\it The NUHM $(\mu, m_A)$ planes for $\tan \beta = 10$, (a) $m_0
= 100$~GeV and $m_{1/2} = 300$~GeV and (b)
$m_0 = 100$~GeV and $m_{1/2} = 500$~GeV, with $A_0 = 0$.
The (red)
dot-dashed lines are the contours $m_h = 114$~GeV, and the near-vertical
(black) dashed lines are the contours $m_{\chi^\pm} = 103.5$~GeV. The
dark (black) dot-dashed lines indicate the GUT stability constraint. Only
the areas inside these curves (small $\mu$) are allowed by this
constraint. The light (turquoise) shaded areas are the cosmologically
preferred regions with
\protect\mbox{$0.1\leq\ohsq\leq 0.3$}. The dark (brick red) shaded
regions is excluded because a charged particle is lighter than the 
neutralino, and the darker (dark blue) shaded regions is excluded because
the LSP is a sneutrino. The medium
(green) shaded region is excluded by $b \to s \gamma$.
The regions allowed by the E821
measurement of $a_\mu$ at the 2-$\sigma$ level
are shaded (pink) and bounded by solid black lines. The solid (blue)
curves correspond to $m_\chi = m_A/2$. }
	\label{muma}
\end{figure}

The NUHM parameter space was recently analyzed \cite{eos3} and a sample
of the results found is shown in Fig. \ref{muma}. While much of the
cosmologically preferred area with $\mu < 0$ is excluded, there is a
significant enhancement in the allowed  parameter space for $\mu > 0$. 

As usual, there are dark (red) regions where there is one or more charged
sparticle lighter than the neutralino $\chi$ so that $\chi$ is no longer the
LSP. First, there are `shark's teeth' at $|\mu| \sim 300$~GeV,
$m_A \la 300$~GeV in panel (a) of Fig.~\ref{muma} where the ${\tilde
\tau}_1$ is the LSP. At small
$|\mu|$, particularly at small $m_A$ when the mass difference $m_2^2 -
m_1^2$ is small, the ${\widetilde \tau}_R$ mass 
is driven small, making the ${\widetilde \tau}_1$ the LSP.  
At even smaller
$|\mu|$, however, the lightest neutralino gets lighter again, 
since $m_\chi \simeq \mu$ when $\mu < M_1 \simeq 0.4 \, 
m_{1/2}$. In addition, there are 
extended dark (red) shaded regions at large $|\mu|$ where left-handed
sleptons become lighter than the neutralino. However, the electron
sneutrino ${\tilde \nu}_e$ and the muon sneutrino ${\tilde \nu}_\mu$
(which are degenerate within the approximations used here) have become
joint LSPs at a slightly smaller $|\mu|$.  Since the possibility of
sneutrino dark matter has been essentially excluded by a combination of
`$\nu$ counting' at LEP, which excludes $m_{\tilde \nu} < 44.7$~GeV
\cite{EFOS}, and searches for cold dark matter, which exclude heavier
${\tilde \nu}$ weighing $\la 1$~TeV \cite{fos}, we still demand
that the LSP be a neutralino $\chi$. The darker (dark blue) shaded
regions are where the sneutrinos are the LSPs, and therefore excluded. 
In the strips adjacent to the
${\widetilde \nu}_{e,\mu}$ LSP regions, neutralino-sneutrino
coannihilation is important in suppressing the relic density to an
acceptable level \cite{eos3}.

The thick cosmological region at smaller $\mu$
corresponds to the `bulk' region familiar from CMSSM studies. The two
(black) crosses indicate the position of the CMSSM points for these input
parameters.  Extending upward in $m_A$ from this region, there is another
light (turquoise) shaded band at smaller $|\mu|$. Here, the neutralino
gets more Higgsino-like and the annihilation to $W^+ W^-$ becomes
important, yielding a relic density in the allowed range~\footnote{This is
similar to the focus-point region \cite{Feng:2000gh} in the CMSSM.}. For
smaller
$|\mu|$, the relic density becomes too small due to the $\chi -
\chi^{\prime} - \chi^+$ coannihilations. For even smaller $|\mu|$ ($
\la 30$~GeV) many channels are kinematically unavailable
and we are no longer near the $h$ and $Z$ pole. As a result the relic
density may again come into the cosmologically preferred region. However,
this region is excluded by the LEP limit on the chargino mass.

The unshaded regions between the allowed bands have a relic density that
is too high: $\Omega_\chi h^2 > 0.3$. However, the $\tilde \tau$ and
${\widetilde \nu}$ coannihilation and bulk bands are connected by
horizontal bands of acceptable relic density that are themselves
separated by unshaded regions of low relic density, threaded by solid
(blue) lines asymptoting to $m_A
\sim 250$~GeV. These lines correspond to cases when $m_\chi \simeq m_A /
2$, where direct-channel annihilation: $\chi + \chi \to A,H$ is important,
and suppresses the relic density \cite{EFGOSi,funnel} creating
`funnel'-like regions.
  
In panel (b), a larger value of $m_{1/2}$ is chosen.
In this case, the previous region excluded by the neutral LSP constraint
at large $|\mu|$, migrates to larger $|\mu|$. The `shark's
teeth' for moderate $|\mu|$ grow, reaching up to $m_A \sim 800$~GeV.
These arise when one combines a large value of $m_{1/2}$ with a
relatively small value of $m_0$, and one may find a $\stau$ or even a
$\tilde e$ LSP.  The large value of $m_{1/2}$ also keeps the rate of $b
\to s \gamma$ under control unless $m_A$ is small.  The chargino
constraint is similar to that in panel (a), whereas the $m_h$ constraint
is irrelevant due to the large value of $m_{1/2}$. Finally, the GUT
mass-squared positivity constraint now allows larger values of $|\mu| \la
1000$~GeV.  

\begin{figure}
\vskip 0.5in
\begin{minipage}{8in}
\epsfig{file=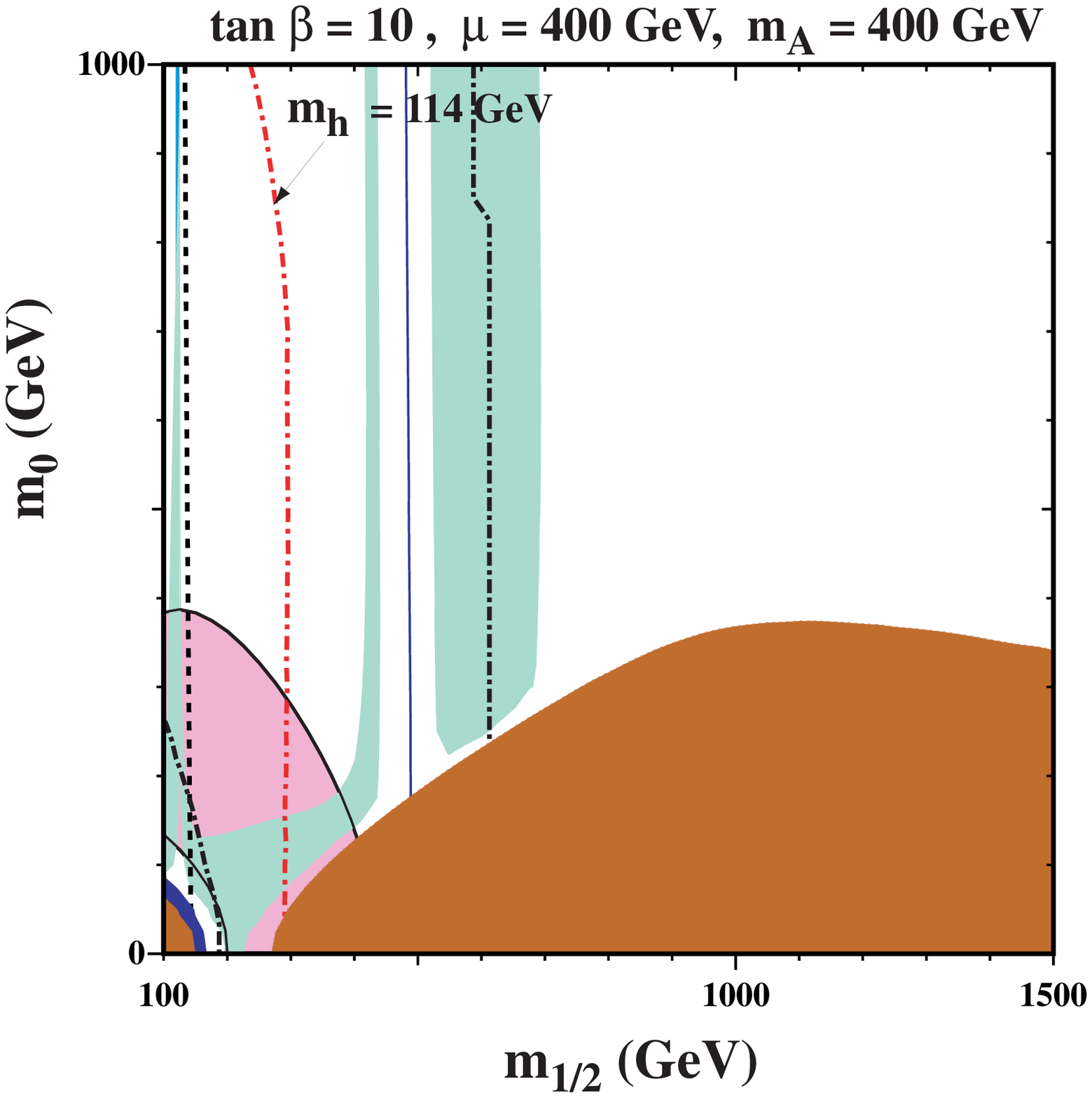,height=3.2in}
\hspace*{-0.17in}
\epsfig{file=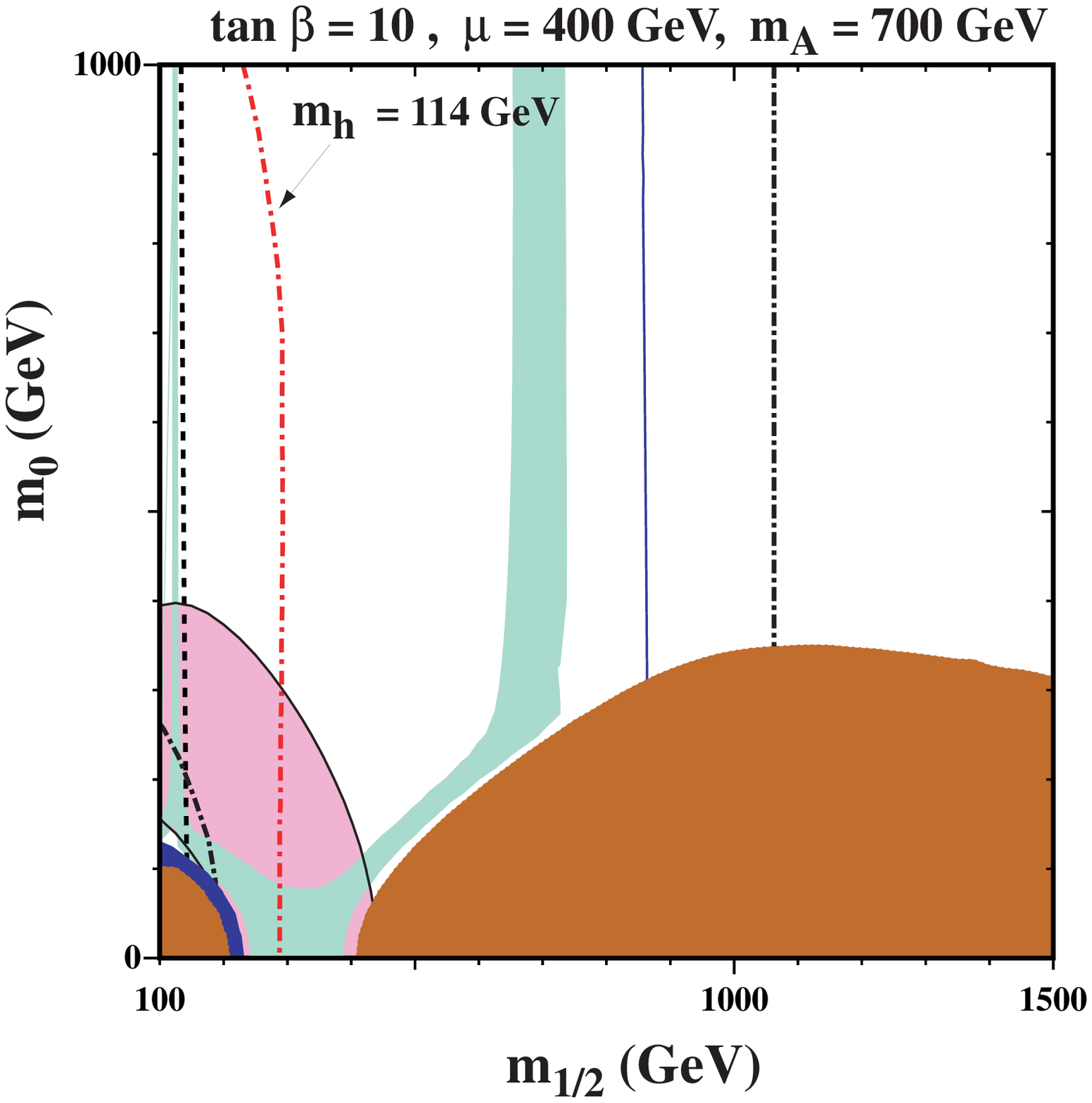,height=3.2in} \hfill
\end{minipage}
\caption{
{\it
Compilations of phenomenological constraints on the MSSM with NUHM
in the $(m_{1/2}, m_0)$ plane for $\tan \beta = 10$ and (a) $\mu = 
400$~GeV, $m_A = 400$~GeV, (b) $\mu = 400$~GeV,
$m_A = 700$~GeV, again assuming $A_0 = 0$.
The shading is consistent with previous figures. 
The (blue) solid line is the contour $m_\chi = m_A/2$, near which
rapid direct-channel annihilation suppresses the relic density.
The dark (black) dot-dashed lines indicates when one or another Higgs
mass-squared becomes negative at the GUT scale: only intermediate values 
of $m_{1/2}$ are allowed. 
}}
\label{fig:m12m0}
\end{figure}

It is also interesting to examine the $m_0$ - $m_{1/2}$ plane in the NUHM
\cite{eos3}. In Fig. \ref{fig:m12m0}a, there is a bulk region at $m_{1/2}
\sim 50$~GeV to 350~GeV, $m_0 \sim 50$~GeV to 150~GeV. 
  As in the CMSSM, the ${\tilde \tau_1}$ is the LSP in the bigger area at
larger $m_{1/2}$, and there are light (turquoise) shaded strips close to
these forbidden regions where coannihilation suppresses the relic density
sufficiently to be cosmologically acceptable. Further away from these
regions, the relic density is generally too high. However, for larger
$m_{1/2}$ there is another suppression, discussed below, which makes the
relic density too low.  At small $m_{1/2}$ and $m_0$ the left handed
sleptons, and also the sneutrinos, become lighter than the neutralino. The
darker (dark blue) shaded area is where a sneutrino is the LSP. 

The near-vertical dark (black) dashed and light (red) dot-dashed lines in
Fig.~\ref{fig:m12m0} are the LEP exclusion contours $m_{\chi^\pm} >
104$~GeV and $m_h > 114$~GeV respectively. As in the CMSSM case, they
exclude low values of
$m_{1/2}$, and hence rule out rapid relic annihilation via direct-channel
$h$ and $Z^0$ poles. The solid lines curved around small values of
$m_{1/2}$ and $m_0$ bound the light (pink) shaded region favoured by
$a_\mu$ and recent analyses of the $e^+ e^-$ data.

A striking feature in Fig.~\ref{fig:m12m0}(a) when $m_{1/2} \sim 500$~GeV
is a strip with low $\ohsq$, which has bands with acceptable relic
density on either side.  The low-$\ohsq$ strip is due to rapid
annihilation via the direct-channel $A, H$ poles which occur when $m_\chi
= m_A / 2 = 200$~GeV, indicated by the near-vertical solid (blue) line.
The analogous rapid-annihilation strips occur in
the CMSSM but at larger $\tan \beta$ as seen in
Fig.~\ref{rd2c50}. There, they are diagonal in the $(m_{1/2}, m_0)$
plane, reflecting a CMSSM link between $m_0$ and $m_A$ that is absent in
this implementation of the NUHM. The right-hand band in
Fig.~\ref{fig:m12m0}(a) with acceptable $\ohsq$ is broadened because the
neutralino acquires significant Higgsino content, and the relic density
is suppressed by the increased $W^+ W^-$ production. As
$m_{1/2}$ increases, the neutralino  becomes almost degenerate with the
second lightest neutralino and the lighter chargino, and the $\chi
- \chi^\prime - \chi^\pm$  coannihilation processes
eventually push $\ohsq < 0.1$ when $m_{1/2} \ga 700$~GeV.   

The two dark (black) dash-dotted lines in Fig.~\ref{fig:m12m0}(a) 
indicate where scalar squared masses become negative at the input GUT
scale for one of the Higgs multiplets.  One of these GUT
stability lines is
near-vertical at $m_{1/2} \sim 600$~GeV, and the other is a curved line at
$m_{1/2} \sim 150$~GeV, $m_0 \sim 200$~GeV.

Panel (b) of Fig.~\ref{fig:m12m0} is for $\mu = 400$~GeV and $m_A =
700$~GeV. We notice immediately that the heavy Higgs pole and the
right-hand boundary of the GUT stability region move out to larger
$m_{1/2} \sim 850, 1050$~GeV, respectively, as one would expect for
larger $m_A$. At this value of $m_A$,  the right side of the
rapid annihilation (`funnel') strip has disappeared, due to enhanced
chargino-neutralino coannihilation effects.

These two examples serve to demonstrate that the $(m_{1/2}, m_0)$ plane
may look rather different in the CMSSM from its appearance in the CMSSM
for the same value of $\tan \beta$. In particular, the locations of
rapid-annihilation funnels and $\chi - \chi^\prime - \chi^\pm$
coannihilation regions are quite model-dependent, and the GUT stability
requirement may exclude large parts of the $(m_{1/2}, m_0)$ plane.

\subsection{Detection}

Because the LSP as dark matter is present locally, there are many
avenues for pursuing dark matter detection. Direct detection techniques
rely on an ample neutralino-nucleon scattering cross-section.
The effective four-fermion lagrangian can be written as
\begin{eqnarray}
\mathcal{L}  & =  &\bar{\chi} \gamma^\mu \gamma^5 \chi \bar{q_{i}} 
\gamma_{\mu} (\alpha_{1i} + \alpha_{2i} \gamma^{5}) q_{i}  \nonumber \\
& + & \alpha_{3i} \bar{\chi} \chi \bar{q_{i}} q_{i} + 
\alpha_{4i} \bar{\chi} \gamma^{5} \chi \bar{q_{i}} \gamma^{5} q_{i} \nonumber \\
& + &\alpha_{5i} \bar{\chi} \chi \bar{q_{i}} \gamma^{5} q_{i} +
\alpha_{6i} \bar{\chi} \gamma^{5} \chi \bar{q_{i}} q_{i} 
\end{eqnarray}
However, the terms involving $\alpha_{1i}, \alpha_{4i}, \alpha_{5i}
$, and
$\alpha_{6i}$  lead to velocity dependent  elastic cross sections.
The remaining terms are: the spin dependent coefficient,
$\alpha_{2i} $  and the scalar coefficient $\alpha_{3i} $.
Contributions to $\alpha_{2i} $ are predominantly through light squark exchange. 	
This is the dominant channel for binos.
Scattering also occurs through Z exchange but this channel 
requires a strong Higgsino component.
Contributions to $\alpha_{3i} $  are also dominated by
light squark exchange 	
but Higgs exchange
is non-negligible in most cases.

\begin{figure}
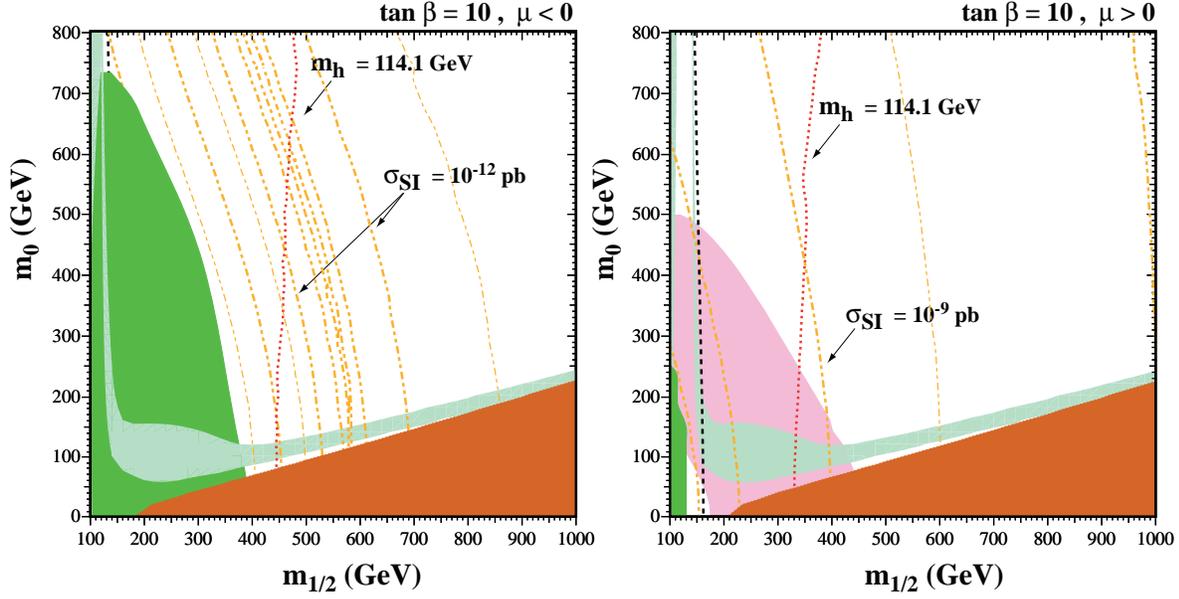

\begin{minipage}{8in}
\epsfig{file=si10n.epss,height=3.2in}
\hspace*{-0.17in}
\epsfig{file=si10p.epss,height=3.2in} \hfill
\end{minipage}
\caption{\label{fig:sicontours}
{\it Spin-independent cross sections in the $(m_{1/2}, m_0)$ planes for
(a) $\tan \beta = 10, \mu < 0$,  (b) $\tan \beta = 10, \mu > 0$.   The
double dot-dashed (orange) curves are contours of the spin-independent
cross section, differing by factors of 10 (bolder) and interpolating
factors of 3 (finer - when shown). For example, in (b), the curves to the
right of the one marked $10^{-9}$ pb correspond to $3
\times  10^{-10}$~pb and $10^{-10}$~pb. 
}}
\end{figure}

Fig.~\ref{fig:sicontours} displays contours of the
spin-independent cross section for the elastic scattering of the LSP
$\chi$ on protons in the $m_{1/2}, m_0$ planes for (a) $\tan \beta = 10,
\mu < 0$, (b) $\tan \beta = 10, \mu > 0$ \cite{eflo3}. The double
dot-dashed (orange) lines are contours of the spin-independent cross
section, and the contours
$\sigma_{SI} = 10^{-9}$~pb in panel (a) and $\sigma_{SI} =
10^{-12}$~pb in panel (b) are
indicated.
The LEP lower limits on $m_h$ and $m_{\chi^\pm}$, as well as the
experimental measurement of $b \to s \gamma$ for $\mu < 0$, tend to bound
the cross sections from above, as discussed in more
detail below. Generally speaking, the spin-independent cross section is 
relatively large in the `bulk' region, but falls off in the coannihilation
`tail'.   Also, we note also that there is a strong
cancellation in the spin-independent cross section when
$\mu < 0$~\cite{EFlO1,EFlO2}, as seen along strips in panel (a)  of
Fig.~\ref{fig:sicontours} where
$m_{1/2} \sim 500$~GeV. In the cancellation region,
the cross section drops lower than $10^{-14}$ pb. All these possibilities
for suppressed spin-independent cross sections are disfavoured by the
data on
$g_\mu - 2$, which favour values of $m_{1/2}$ and
$m_0$ that are not very large, as well as $\mu > 0$, as seen in panel
(b) of Fig.~\ref{fig:sicontours}. Thus $g_\mu - 2$ tends to provide a
lower bound on the spin-independent cross section.

\begin{figure}
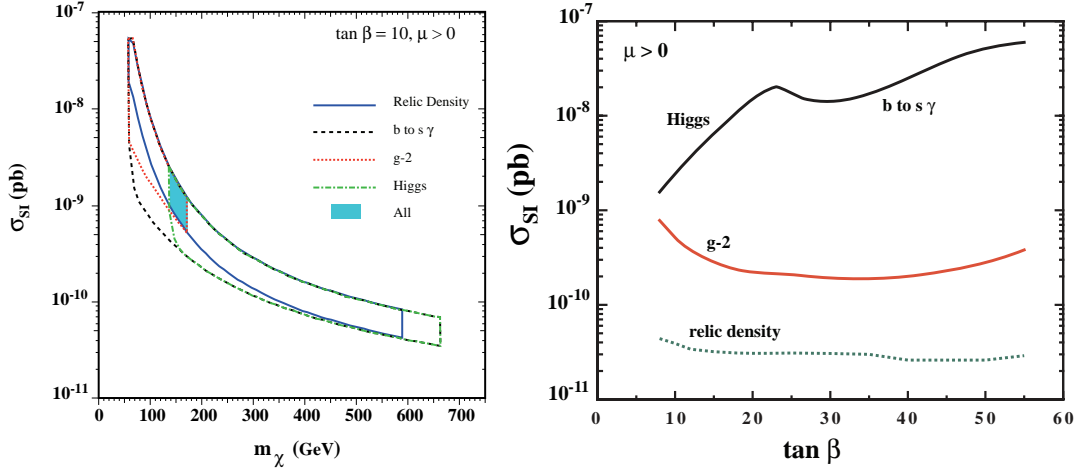

\begin{minipage}{6in}
\begin{center}
\epsfig{file=sirange10p.epss,height=2.5in}
\epsfig{file=sirangep.epss,height=2.5in} \hfill
\end{center}
\end{minipage}
\caption{\label{fig:decimation}
{\it Allowed ranges of the cross sections for $\tan \beta = 10$ (a)
$\mu > 0$ for spin-independent elastic scattering.  The solid (blue)
lines  indicate the
relic density constraint, the dashed (black) lines the $b
\to s \gamma$ constraint, the dot-dashed (green) lines the
$m_h$ constraint, and the dotted (red) lines the $g_\mu -
2$ constraint. The shaded (pale blue)  region is allowed by all
the constraints. (b) The allowed ranges of the spin-independent cross
section for  $\mu > 0$. The darker solid (black) lines show the upper 
limits on 
the cross sections obtained from $m_h$ and $b \to s \gamma$, and (where 
applicable) the lighter 
solid (red) lines show the lower limits suggested by $g_\mu - 2$ and the 
dotted (green) lines the lower limits from the relic density.
}}
\end{figure}

Fig.~\ref{fig:decimation}(a) illustrates the effect on the cross sections
of each of the principal phenomenological constraints, for the particular
case $\tan \beta = 10$  $\mu > 0$. The solid
(blue) lines mark the bounds on the cross sections allowed by the
relic-density constraint $0.1 < \Omega_\chi h^2 < 0.3$
alone. 
For any given value of
$m_{1/2}$, only a restricted range of $m_0$ is allowed. Therefore, only a
limited range of $m_0$, and hence only a limited range for the cross
section, is allowed for any given value of
$m_\chi$. The thicknesses of the allowed regions are due in part to the 
assumed
uncertainties in the nuclear inputs.  These have been discussed at
length in \cite{EFlO2,EFlO1}.
 On 
the other hand, a broad range of $m_\chi$ is allowed, when one takes into
account the coannihilation `tail' region at each $\tan 
\beta$ and the rapid-annihilation `funnel' regions for $\tan
\beta = 35, 50$. The dashed (black) line  displays the range allowed by
the $b \to s
\gamma$ constraint alone. In
this case, a broader range of $m_0$ and hence the spin-independent cross
section is possible for any given value of $m_\chi$. The impact of the
constraint due to $m_h$ is shown by the dot-dashed (green) line. 
Comparing with the previous constraints, we see that a region at low
$m_\chi$ is excluded by $m_h$, strengthening significantly the previous
{\it upper} limit on the spin-independent cross section. Finally, the
dotted (red)  lines in Fig.~\ref{fig:decimation} show the impact of the
$g_\mu - 2$ constraint. This imposes an upper bound on
$m_{1/2}$ and hence $m_\chi$, and correspondingly a {\it lower} limit on
the spin-independent cross section.

This analysis is 
extended 
in panel (b)  of Fig.~\ref{fig:decimation} to all the values $8 
< \tan \beta \le 55$ and we find overall that \cite{eflo3}
\begin{eqnarray}
2 \times 10^{-10}~{\rm pb} \la & \sigma_{SI} & \la 6 \times 10^{-8}~{\rm
pb},
\\ 2 \times 10^{-7}~{\rm pb} \la & \sigma_{SD} & \la  10^{-5}~{\rm
pb},
\label{xsecrangeallp}
\end{eqnarray}
for $\mu > 0$. ($\sigma_{SD}$ is the spin-dependent cross-section not
shown in the figures presented here.) As we see in panel (b)  of
Fig.~\ref{fig:decimation},
$m_h$ provides the most  important upper limit 
on the cross sections for $\tan \beta < 23$, and $b \to s \gamma$ for 
larger $\tan \beta$, with $g_\mu - 2$ always providing a more stringent 
lower limit than the relic-density constraint. 
The relic density constraint shown is evaluated at the endpoint of the
coannihilation region. At large $\tan \beta$,  the
Higgs funnels or the focus-point regions have not been considered, as
their locations are very  sensitive to input parameters and calculational
details~\cite{EO}.

\begin{figure}[hbtp]
	\centering 
\hspace*{1.30in}
\begin{minipage}{8in}
\epsfig{file=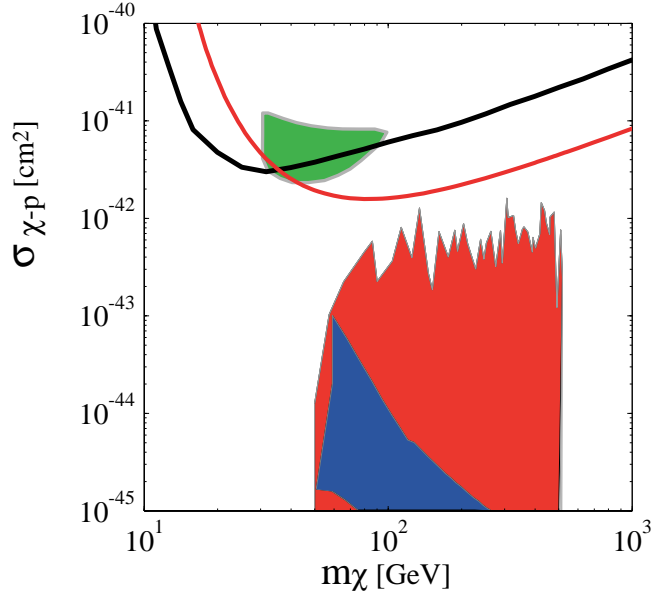,height=3.2in}
\end{minipage}
	\caption{\it Limits from the CDMS \protect\cite{cdms} and
Edelweiss
\protect\cite{edel} experiments on the neutralino-proton
elastic scattering cross section as a function of the neutralino mass. The
Edelweiss limit is stronger at higher $m_\chi$. These results nearly
exclude the shaded region observed by DAMA \protect\cite{dama}. The
theoretical predictions lie at lower values of the cross section.}
	\label{cdms}
\end{figure}

The results from a CMSSM and MSSM analysis \cite{EFlO1,EFlO2} for $\tan
\beta = 3$ and 10 are compared with the most recent CDMS \cite{cdms} and
Edelweiss \cite{edel} bounds in Fig.~\ref{cdms}. These results have nearly
entirely excluded the region purported by the DAMA \cite{dama} experiment.
The CMSSM prediction
\cite{EFlO1} is shown  by the dark shaded region, while the NUHM case \cite{EFlO2} is
shown by the larger lighter shaded region.

\begin{figure}
\centering
		\epsfxsize=10cm
\epsfbox{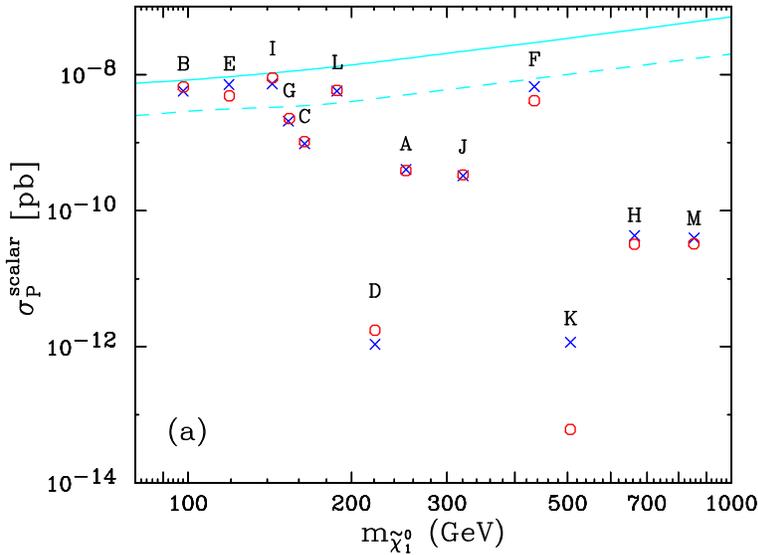}
\caption{\it Elastic spin-independent scattering  
of supersymmetric relics on protons  calculated in 
benchmark scenarios\protect\cite{EFFMO}, compared with the 
projected sensitivities for CDMS
II~\protect\cite{Schnee:1998gf} and CRESST \protect\cite{Bravin:1999fc}
(solid) and GENIUS \protect\cite{GENIUS} (dashed).
The predictions of our code (blue
crosses) and {\tt Neutdriver} \protect\cite{neut} (red circles) for
neutralino-nucleon scattering are compared.
The labels correspond to the benchmark points as shown in 
Fig.~\protect\ref{fig:rough}.}
\label{fig:DM}
\end{figure}

I conclude by showing the
prospects for direct detection for the benchmark points discussed
above\cite{EFFMO}. Fig.~\ref{fig:DM} shows rates for the elastic
spin-independent scattering of supersymmetric relics,
including the projected sensitivities for CDMS
II~\cite{Schnee:1998gf} and CRESST \cite{Bravin:1999fc} (solid) and
GENIUS \cite{GENIUS} (dashed).
Also shown are the cross sections 
calculated in the proposed benchmark scenarios discussed in the previous
section, which are considerably below the DAMA \cite{dama} range
($10^{-5} - 10^{-6}$~pb).
Indirect searches for supersymmetric dark matter via the products of
annihilations in the galactic halo or inside the Sun also have prospects
in some of the benchmark scenarios \cite{EFFMO}.

\section*{Acknowledgments}
I would like to thank J. Ellis, T. Falk, A. Ferstl, G. Ganis, Y. Santoso, and  M.
Srednicki for enjoyable collaborations from which this work is culled.
 This work was supported in part by 
DOE grant DE-FG02-94ER40823 at Minnesota.

\end{document}